%% file: eprint.tex
\pdfoutput=1

\documentclass[10pt]{article}
\usepackage{graphicx}

\usepackage{lineno}

\usepackage[utf8]{inputenc}

\def\Title#1{\begin{center} {\Large #1 } \end{center}}
\def\Author#1{\begin{center}{ \sc #1} \end{center}}
\def\Address#1{\begin{center}{ \it #1} \end{center}}

\newcommand\pubblock{\rightline{\begin{tabular}{l} Proceedings of the Fifth Annual LHCP\\ \pubnumber\\
         \pubdate  \end{tabular}}}

\newenvironment{Abstract}{\begin{quotation} \begin{center} 
             \large ABSTRACT \end{center}\bigskip 
      \begin{center}\begin{large}}{\end{large}\end{center} \end{quotation}}

\newenvironment{Presented}{\begin{quotation} \begin{center} 
             PRESENTED AT\end{center}\bigskip 
      \begin{center}\begin{large}}{\end{large}\end{center} \end{quotation}}

\def\Acknowledgements{\bigskip  \bigskip \begin{center} \begin{large}
             \bf ACKNOWLEDGEMENTS \end{large}\end{center}}

\input econfmacros.tex

\usepackage{color}

\newcommand\pubnumber{ LHCb-PROC-2017-024 }

\newcommand\pubdate{August 8, 2017}

\def\affiliation{
École polytechnique fédérale de Lausanne, 1015 Lausanne, Switzerland}

\begin{document}

\large
\begin{titlepage}
\pubblock

\vfill
\Title{Tests of Lepton Flavour Universality with $b \rightarrow s\ell \ell$ transitions at LHCb}
\vfill

\Author{Guido Andreassi}
\Address{\affiliation}
\begin{center}
On behalf of the LHCb collaboration
\end{center}
\vfill
\begin{Abstract}

Semileptonic $b \rightarrow s\ell \ell$ processes constitute a good probe for new physics phenomena: new particles contributing to the loops could affect branching fractions and angular distributions, and have different couplings to different lepton families, thus violating lepton flavour universality.\\
Recent results from the LHCb experiment are reviewed.

\end{Abstract}
\vfill

\begin{Presented}
The Fifth Annual Conference\\
 on Large Hadron Collider Physics \\
Shanghai Jiao Tong University, Shanghai, China\\ 
May 15-20, 2017
\end{Presented}
\vfill
\end{titlepage}
\def\thefootnote{\fnsymbol{footnote}}
\setcounter{footnote}{0}
%

\normalsize 


\section{Introduction}

\begin{figure}[htb]
\centering
\includegraphics[width=.3\textwidth]{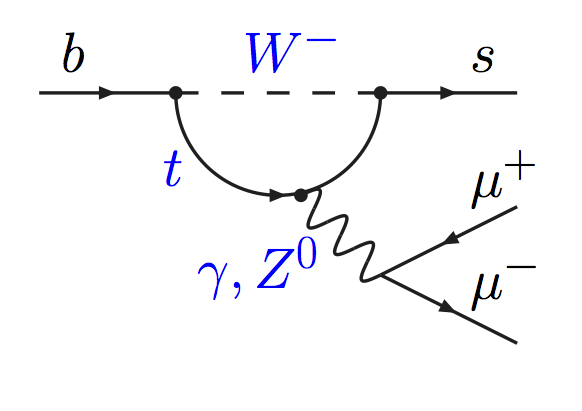}
\includegraphics[width=.3\textwidth]{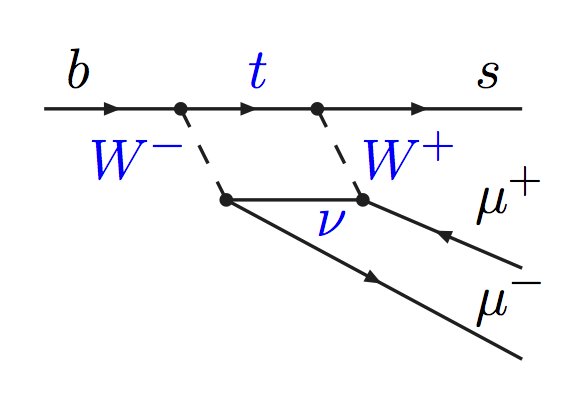}
\caption{Penguin (left) and box (right) Feynman diagrams describing a $b \rightarrow s\ell \ell$ transition.}
\label{fig:feynmann}
\end{figure}

The coupling of the leptons to gauge bosons in the Standard Model (SM) of particle physics is not predicted to depend on the flavour. This property is known as lepton flavour universality (LFU).\\
$b \rightarrow s\ell \ell$ transitions constitute a good probe for new physics searches in general and LFU tests in particular. Such processes are indeed rare in the SM, being forbidden at tree level and only allowed via higher order diagrams such as those shown in Figure~\ref{fig:feynmann}. The presence of new, yet unobserved, particles entering the loops could alter their branching ratios and/or angular distributions.\\
In some theory models like those predicting the existence of leptoquarks \cite{Becirevic:2016yqi, Crivellin:2017zlb} or Z' bosons \cite{Descotes-Genon:2013wba, Gauld:2013qja, Buras:2012dp, Altmannshofer:2013foa, Altmannshofer:2014cfa, Altmannshofer:2016jzy}, new contributions to $b \rightarrow s\ell \ell$ would introduce a violation of LFU.\\

Two results obtained from the analysis of the LHC Run-1 data collected by the LHCb experiment in 2011 and 2012 at a centre-of-mass energy of $\sqrt{7}$ and $\sqrt{8}$ $Tev/c^2$ respectively are presented hereinafter.

\section{Observations}

\subsection{$R_K$}
In 2014 the LHCb collaboration tested LFU using $B^+ \rightarrow K^+ \ell\ell$ decays \cite{Aaij:2014ora}, via the measurement of the ratio
\begin{equation}
R_K = \frac{\int_{q^2_{min}}^{q^2_{max}}\frac{d\Gamma(B^+ \rightarrow K^+ \mu^+ \mu^-)}{dq^2}dq^2}{\int_{q^2_{min}}^{q^2_{max}}\frac{d\Gamma(B^+ \rightarrow K^+ e^+ e^-)}{dq^2}dq^2}
\label{eq:RK}
\end{equation}
in the range $1$ GeV$^2/c^4 < q^2 < 6 $ GeV$^2/c^4$, where $q^2$ is the invariant mass of the dilepton system.\\
Due to LFU, $R_K$ in the SM is predicted to be $1 \pm \mathcal{O}(10^{-3})$ \cite{Bobeth:2007dw, Bordone:2016gaq}.\\

\begin{figure}[htb]
\centering
\includegraphics[width=.5\textwidth]{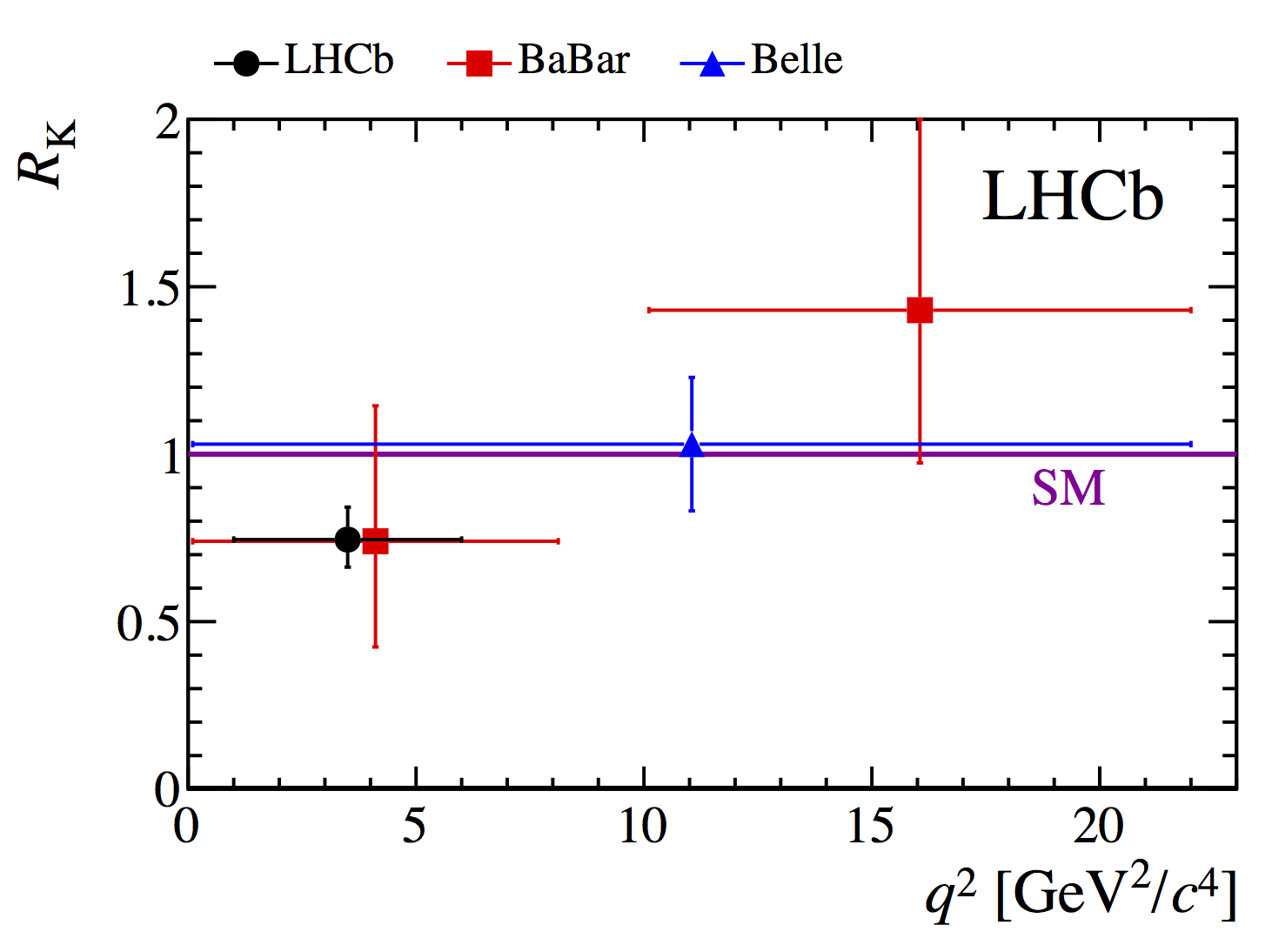}
\caption{Comparison of the measurements of $R_K$ from LHCb (black dots), BaBar \cite{Lees:2012tva} (red squares) and Belle \cite{Wei:2009zv} (blue triangles) with the SM expectation (purple line).}
\label{fig:RK_comparison}
\end{figure}

From the experimental point of view, electrons and muons behave very differently in the LHCb detector. In particular, while the latter are characterised by a high reconstruction efficiency and a very clean signature, the former emit large amounts of bremsstrahlung radiation, which implies a significant degradation of the resolution on the invariant dilepton mass, partially recovered by dedicated algorithms in the reconstruction software. Moreover, different levels of background contamination are present in the two channels, which
implies substantial differences in the analysis. To minimise the effect of systematic uncertainties, at LHCb the measurement has been performed as a double ratio of branching fractions
\begin{equation}
R_K = \frac{\mathcal{B}(B^+ \rightarrow K^+ \mu^+ \mu^-)}{\mathcal{B}(B^+ \rightarrow  K^+ J/\psi (\rightarrow \mu^+ \mu^-))} \Bigg/ \frac{\mathcal{B}(B^+ \rightarrow  K^+ J/\psi (\rightarrow e^+ e^-))}{ \mathcal{B}(B^+ \rightarrow K^+ e^+ e^-)}.
\end{equation}
Candidates for the normalisation channel $B^+ \rightarrow  K^+ J/\psi (\rightarrow \ell^+ \ell^-)$ are selected using the same criteria as the non-resonant counterpart.\\
The signal yields are extracted through a fit to the invariant $B$ mass. In the case of the electron channel, the sample is split in categories depending on whether the event was triggered by the electron, the kaon or any other particle in the event. This allows to treat individually the different efficiencies of these three categories.\\
The measured value of $R_K$ is $0.745^{+0.090}_{-0.074}(stat)^{+0.036}_{-0.036}(syst)$, which is in tension with the SM at $2.6 \sigma$ level. This value is found to be consistent with Ref.~\cite{Aaij:2014pli}. Figure~\ref{fig:RK_comparison} shows a comparison with previous measurements from the B-factories, which are compatible with the SM within one standard deviation.

\subsection{$R_{K^*}$}
A new result from the LHCb experiment is the measurement of $R_{K^{\ast}}$~\cite{Aaij:2017vbb}. Similarly to the ratio in Equation \ref{eq:RK}, $R_{K^{\ast}}$ is defined as
\begin{equation}
R_K = \frac{\int_{q^2_{min}}^{q^2_{max}}\frac{d\Gamma(B^0 \rightarrow K^{\ast 0} \mu^+ \mu^-)}{dq^2}dq^2}{\int_{q^2_{min}}^{q^2_{max}}\frac{d\Gamma(B^0 \rightarrow  K^{\ast 0} e^+ e^-)}{dq^2}dq^2}
\label{eq:RK*}
\end{equation}
with $K^{\ast 0} \rightarrow K^+ \pi^-$.
As in the $R_K$ analysis, the differences in the selection of muons and electrons constitute a challenge and, in order to reduce the systematic errors, the measurement is performed as a double ratio
\begin{equation}
R_K = \frac{\mathcal{B}(B^0 \rightarrow K^{\ast 0} \mu^+ \mu^-)}{\mathcal{B}(B^0 \rightarrow  K^{\ast 0} J/\psi (\rightarrow \mu^+ \mu^-))} \Bigg/ \frac{\mathcal{B}(B^0 \rightarrow  K^{\ast 0} J/\psi (\rightarrow e^+ e^-))}{ \mathcal{B}(B^0 \rightarrow K^{\ast 0} e^+ e^-)}.
\end{equation}
The analysis is performed in the two $q^2$ bins $0.045$ GeV$^2/c^4 < q^2 < 1.1$ GeV$^2/c^4$ and $1.1$ GeV$^2/c^4 < q^2 < 6$ GeV$^2/c^4$, referred to as \textit{low} and \textit{central} $q^2$ region, respectively. The thresholds are chosen to match the dimuon kinematic threshold, to include the $\Phi(1020) \rightarrow \ell \ell$ contribution in the first bin and to reduce contamination from the radiative tail of the $J/\psi$ resonance.\\

The signal in the two $q^2$ bins both in the $ee$ and $\mu\mu$ case are fitted simultaneously with the corresponding resonant counterpart. For the electron channel, similarly to the  $R_K$ analysis, the sample is further divided into three trigger categories: candidates for which one of the electrons satisfies the hardware electron trigger, candidates for which one of the hadrons from the $K^{\ast}$ decay meets the hardware hadron trigger requirements, and candidates triggered by activity in the event not associated with any of the signal decay particles.\\

Several different cross-checks have been performed to ensure the absence of biases in the analysis. The most stringent one is the measurement of the ratio
\begin{equation}
r_{J/\psi} = \frac{\mathcal{B}(B^0 \rightarrow  K^{\ast 0} J/\psi (\rightarrow \mu^+ \mu^-))}{\mathcal{B}(B^0 \rightarrow  K^{\ast 0} J/\psi (\rightarrow e^+ e^-))}
\end{equation}
which is expected to be $1$ due to the domination of the SM contribution from the $J/\psi$ resonance. $r_{J/\psi}$ is found to be consistent with unity and constant with respect to the kinematics of the decay.\\

The measured values of $R_{K^{\ast}}$ are reported in Table \ref{tab:RK*} along with their compatibility with the SM.\\
The results are found to be in good agreement between the three trigger categories. Furthermore, the branching fraction of the decay $B^0 \rightarrow K^{\ast 0} \mu^+ \mu^-$ is also measured and found to be in good agreement with Ref.~\cite{Aaij:2016flj}.

\begin{figure}[htb]
\centering
\includegraphics[width=\textwidth]{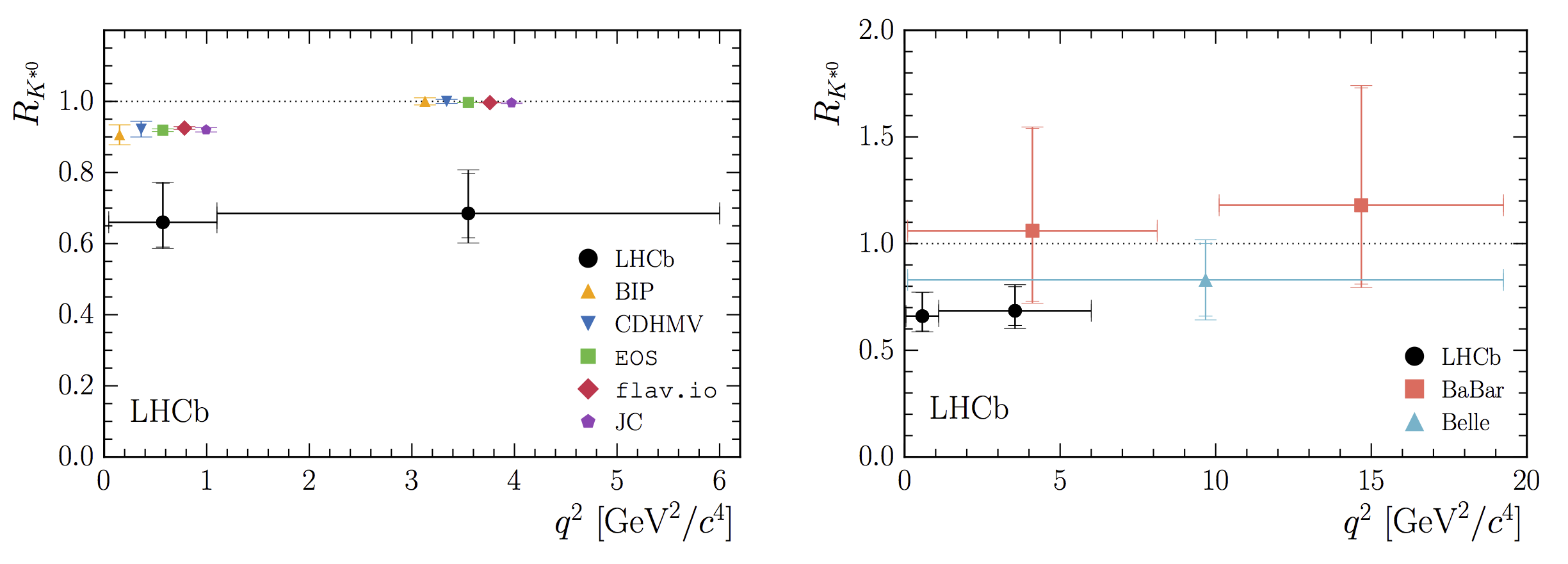}
\caption{Comparison of the measurements of $R_{K^{\ast}}$ from LHCb with (left) SM predictions and (right) BaBar \cite{Lees:2012tva} and Belle \cite{Wei:2009zv}.}
\label{fig:RK*_comparison}
\end{figure}

Comparisons with previous measurements from the B-factories and theory predictions are shown in Figure~\ref{fig:RK*_comparison}.\\

\begin{table}[t]
\begin{center}
\begin{tabular}{l|ccc}  
$q^2$ bin &  $R{K^{\ast}}$ & compatibility with SM\\
\hline\\
 low  &   $0.66 ^{+0.11}_{-0.07} \pm 0.03$     & $2.1-2.3 \, \sigma$ \\[8pt]
 central & $0.69 ^{+0.11}_{-0.07} \pm 0.05$ & $2.4-2.5 \, \sigma$
\end{tabular}
\caption{$R{K^{\ast}}$ measured values in the two $q^2$ intervals taken into account, along with their compatibility with the SM. The spread in the significances is due to small differences in the theory predictions from the models considered. The first uncertainties are statistical and the second are systematic.}
\label{tab:RK*}
\end{center}
\end{table}

\section{Interpretations}
Both $R_K$ and $R_{K{\ast}}$ measurements have raised interest in the world of particle physics theory and phenomenology.  Several global fits taking into account observables from different experiments, including results from LFU searches and $b \rightarrow s\ell \ell$ angular analyses, have been performed in an attempt to interpret these anomalies by constraining new physics contributions in Wilson coefficients~\cite{Descotes-Genon:2015uva, Altmannshofer:2017fio, Mahmoudi:2016mgr, Aaij:2017vbb}. Links are also predicted to exist between lepton flavour universality and lepton flavour violation~\cite{Hiller:2016kry, Glashow:2014iga} and investigations in this direction can help build a consistent picture.\\
Further effort is still required both from the experimental and theoretical side to shed a clear light on the present scenario.

\section{Conclusions}
Analyses of $b \rightarrow s \ell \ell$ decays are providing interesting and consistent hints of lepton flavour universality violation, which deserve further investigation. The upcoming results from the LHC Run-2 data, along with new studies on lepton flavour universality and lepton flavour violation will certainly lead towards clearer conclusions.

\Acknowledgements
I am very grateful to Simone Bifani, Luca Pescatore and Martino Borsato for their support in the preparation of the material that I presented at LHCp 2017.

\end{document}

%% file: econfmacros.tex
\def\beq{\begin{equation}}
\def\eeq#1{\label{#1}\end{equation}}
\def\eeqn{\end{equation}}

\def\beqa{\begin{eqnarray}}
\def\eeqa#1{\label{#1}\end{eqnarray}}
\def\eeqan{\end{eqnarray}}

\let\bar=\overbar

\def\Dslash{\not{\hbox{\kern-4pt $D$}}}
\def\dslash{\not{\hbox{\kern-2pt $\del$}}}

\def\msb{{\bar{\ssstyle M \kern -1pt S}}}

